\documentclass[prl,twocolumn,floats,aps,showpacs,nofootinbib,amssymb]{revtex4}
\usepackage{subfigure}
\usepackage[dvips]{graphicx}
\usepackage{amsmath}
\usepackage{bm}
\usepackage{times}
\usepackage{epsfig}
\newcommand{\nonubb}{\mbox{$ 0\nu\beta\beta $ }}

\newcommand{\be}{\begin{equation}}
\newcommand{\ee}{\end{equation}}

\begin{document}
\title{$0 \nu\, \beta \beta$ decay process in left-right symmetric models without scalar Bidoublet}
\author{\bf Sudhanwa Patra}
\email{sudha.astro@gmail.com}
\affiliation{Center of Excellence in Theoretical and Mathematical Sciences, \\
Siksha 'O' Anusandhan University, Bhubaneswar-751030, India}

\begin{abstract}

We present an alternative formulation of the left-right symmetric theory where 
the scalar sector consists of two Higgs doublets which differs from the standard 
version of the left-right model that makes use of $L-$ and $R-$ Higgs triplets and 
a Higgs bi-doublet. The basic idea is to consider few extra charged iso-singlet fields 
and the fermion masses can be realized by integrating out these heavy isosinglet fields.
We also give a detailed discussion on neutrinoless double beta decay in this 
particular left-right symmetric theory where the right-handed Majorana neutrino 
can be of MeV range. With this right-handed Majorana mass around MeV scale, the contribution 
to neutrinoless double beta decay coming from the right-handed current can be 
comparable with the contributions coming from the standard left-handed sector only if
the right-handed gauge boson mass is around 5 TeV, and with this operative scale of $W_R$ 
around few TeV, it is possible to probe at LHC. We have briefly commented on cosmological 
constraints coming from the big-bang nucleosynthesis and Universe cosmology to the 
right-handed neutrinos involved in this discussion.

\end{abstract}

\maketitle

\section{Introduction}
Left-Right symmetric model(LRSM) is a novel extension of the standard model of 
particle physics, which will treat the left-handed and right-handed particles 
on equal footing, and the parity violation we observe at low energies would be 
due to the spontaneous breaking of the left-right symmetry at some high scale 
\cite{Mohapatra:1974gc, Senjanovic:1975rk, Mohapatra:1974hk, Mohapatra:1980qe, 
Mohapatra:1979ia, Deshpande:1990ip}. The right handed neutrino 
is an automatic consequences of left-right symmetric theory, such models provide 
a natural explanation for the smallness of neutrino masses via see-saw mechanism \cite{Minkowski:1977sc, 
Yanagida:1979as, Mohapatra:1979ia}. Another interesting feature of 
the left-right symmetric model is that the difference between the baryon number 
($B$) and the lepton number ($L$) becomes a gauge symmetry, which leads to several 
interesting consequences.

However, till now, the fundamental fermionic representation and the Higgs sector 
is not fully determined which in turn gives a variety of possibilities of choosing 
these representation (of course the representations has to be restricted by the 
symmetry of the known gauge group). In addition. one has to address the issue of origin 
of the observed fermion masses and mixing. In the standard model (SM), all the flavor 
structure is determined by unknown Yukawa couplings. Hence, a new approach to address these issues has been discussed 
in  Ref. \cite{Berezhiani:1983hm, Berezhiani:1985in, Berezhiani:1991ds, Berezhiani:1992pj}. 
The basic structure of these models excludes the conventional Higgs triplets and bidoublet, 
but includes the new left-handed Higgs doublet $\Phi_L$ and the right-handed Higgs doublet 
$\Phi_R$ and the masses of the usual fermions can be realized by means of a universal 
seesaw with the aid of few extra isosinglet fermions. In this paper, we shall follow a 
simplest approach which contains scalar sector with only two Higgs doublets and few 
extra iso-singlet fermions in order to realize fermion masses and mixings. 

Furthermore, the experimental observation on solar, atmosphere, reactor and accelerator 
neutrino oscillations have revealed that neutrinos can oscillate from one flavor 
to another as they propagate is the strongest indication for nonzero neutrino masses 
and mixing \cite{Fukuda:2001nk, Ahmad:2002jz, Ahmad:2002ka, Bahcall:2004mz}. 
Moreover,until now there is no information about the 
absolute scale of neutrino masses. One can find the bound on absolute scale of neutrino mass 
via studies of lepton number ($L$) violating neutrino less double $\beta$-decay 
($^{A}_{Z}\left[\text{Nucl}\right] \rightarrow \;^{\;\;\;\;A}_{Z+2}\left[\text{Nucl}' 
\right] + 2 e^-$), whose observation would imply that neutrinos are Majorana fermions 
\cite{Weinberg:1979sa}. At present days, the best limit on the half life of this process is  
$T_{1/2} < 3 \times 10^{25}$ years coming from the Heidelberg-Moscow \cite{KlapdorKleingrothaus:2000sn, 
KlapdorKleingrothaus:2006bd} and IGEX \cite{Aalseth:2002rf}, collaborations conducted experiments with 
$^{76}\text{Ge}$, \cite{KlapdorKleingrothaus:2004wj} 
which in turn translated to a bound on the effective neutrino mass  
$m_{\mathrm{eff}} \leq 0.21 - 0.53~ {\rm eV} $, 
where the maximum and minimum range arises due to the uncertainty in the nuclear matrix elements. 
In addition to this bound, there are other upcoming experiments trying to improve this bound \cite{Arnaboldi:2008ds, 
Avignone:2007fu, GomezCadenas:2011it}.

Along with the standard contribution to $\nonubb$ which comes through the exchange of light 
neutrinos (where the effective Majorana neutrino mass is just the absolute value of the ($ee$) 
element of the low energy neutrino mass matrix in the flavour basis), there can be many other 
contribution to neutrinoless double beta in generic left-right (LR) models \cite{Pascoli:2007qh,Hirsch:1995ek, 
Pas:1998nn}. The importance of RH Majorana neutrinos for neutrinoless double beta decay has been 
pointed out by Mohapatra \cite{Mohapatra:1986su} while Doi and Kotani \cite{Doi:1992dm} gave a detailed 
discussion of decay rate including terms for both left-handed and right-handed Majorana neutrinos. 
Recently, a very interesting possibility of ``left-right symmetry: from LHC to 
neutrinoless double beta decay'' \cite{Tello:2010am, Nemevsek:2011aa} has been proposed, wherein the scale of 
left-right symmetry restoration and associated lepton number violation (the neutrinoless 
double beta decay) can be probed at LHC. Then this idea has been discussed in great detailed 
in Ref. \cite{Chakrabortty:2012mh} where the scale of new physics is at $\sim$ TeV scale which is 
phenomenologically rich for LHC.

Since the aforementioned LR symmetric model without bidoublet offers an appealing 
possibility that both the light and heavy Majorana neutrino mass matrices are related 
with each other and hence, it will be worth to study the neutrinoless double decay 
process in this scenario including both the contributions coming from left-handed as 
well as right-handed sector. With this motivation, we shall first present the LR models 
with only isodoublets Higgs $\Phi_L$ and $\Phi_R$ without having a scalar bidoublet with 
detailed discussion. We then extend our discussion to $\nonubb$ with particular emphasis 
on new contribution coming from right-handed current.
\section{The model}
We now recapitulate the important features of the minimal left-right symmetric model 
without any scalar bidoublet where spontaneous parity breaking occurs through only Higgs 
doublets which has been discussed in Ref. \cite{Berezhiani:1983hm, Berezhiani:1985in, Berezhiani:1991ds, Berezhiani:1992pj,Gu:2010yf}. 
At this stage, we shall write 
the particle content and corresponding Lagrangian for the aforementioned minimal model without 
invoking any horizontal symmetry, although inclusion of horizontal symmetry is more complete one. 
The gauge group of this particular model is $SU(3)_C \times SU(2)_L \times SU(2)_R \times U(1)_{B-L}$, 
where the electric charge is related to the generators of the group as:
\begin{equation}\label{1}
    Q = T_{3L} + T_{3R} + {B-L \over 2} = T_{3L} + Y\,.
\end{equation}

The fermion content of the minimal $SU(3)_C \times SU(2)_L \times SU(2)_R \times U(1)_{B-L}$ 
gauge model is well-known, i.e. quarks and leptons transform under the left-right 
symmetric gauge group as: 
\begin{eqnarray*}
q_{L}=\begin{pmatrix}u_{L}\\
d_{L}\end{pmatrix}\equiv[3,2,1,{\frac{1}{3}}] & , & q_{R}=\begin{pmatrix}u_{R}\\
d_{R}\end{pmatrix}\equiv[3,1,2,{\frac{1}{3}}]\,,\nonumber \\
\ell_{L}=\begin{pmatrix}\nu_{L}\\
e_{L}\end{pmatrix}\equiv[1,2,1,{-1}] & , & \ell_{R}=\begin{pmatrix}N_{R}\\
e_{R}\end{pmatrix}\equiv[1,1,2,-1]
\end{eqnarray*}
In the generic left-right models, a scalar bidoublet transforming $(1, 2, 2, 0)$ is introduced for 
obvious reason that we want masses for quarks and leptons. 
Also there are few attempts 
has been made in order to explain fermion masses in minimal left-right symmetric models without 
adding scalar bidoublet and, in this case, scalar doublets were added to do the job. 

The simplest way to achieve this symmetry breaking is to introduce two Higgs doublets 
which are given below
 \begin{eqnarray}
\Phi_{L}=\left( \phi^{+}_L, \phi^{0}_L \right), \quad 
\Phi_{R}=\left( \phi^{+}_R, \phi^{0}_R \right)
 \label{eq:lrhiggs}
\end{eqnarray}

Thus, the complete Lagrangian density could be read as:
\begin{eqnarray}
L&=&-\frac{1}{4}W_{\mu\nu L}.W^{\mu\nu L}-\frac{1}{4}W_{\mu\nu R}.W^{\mu\nu R}- 
\frac{1}{4}B_{\mu\nu}B^{\mu\nu}\nonumber\\
&+&\overline{\psi}_{L}\, \gamma^{\mu} 
\left(i\partial_{\mu}-g\frac{1}{2}\tau.W_{\mu L}-g'\frac{Y}{2}B_{\mu}\right) 
\psi_{L}\nonumber\\
&+&\overline{\psi}_{R}\,\gamma^{\mu}\left(i\partial_{\mu}-g\frac{1}{2} 
\tau.W_{\mu R}-g'\frac{Y}{2}B_{\mu}\right)\psi_{R}\nonumber\\
&+&\left|\left(i\partial_{\mu}- 
g\frac{1}{2}\tau.W_{\mu L}-g'\frac{Y}{2}B_{\mu}\right)\Phi_{L}\right|^2\nonumber\\
&+& 
\left|\left(i\partial_{\mu}-g\frac{1}{2}\tau.W_{\mu R}-g'\frac{Y}{2}B_{\mu}\right) 
\Phi_{R}\right|^2     \nonumber\\
&-&V(\Phi_{L},\Phi_{R}) 
\label{eq:completelagrangian}
\end{eqnarray}
where $g_{L}=g_{R}=g$ are the $SU(2)$ couplings, $g'$ is the $U(1)$ coupling, 
$\gamma^{\mu}$ are the Dirac matrices, $\tau$'s are the Pauli spin matrices, 
$V(\Phi_{L},\Phi_{R})$ is the Higgs potential, $Y$ is the hypercharge ($Y=B-L$).
Also $\psi$ is a fermionic spinner valid for both quarks ($q$) and leptons ($\ell$).
Here, the vacuum expectation values of two doublets ($v_L$ and 
$v_R$ with the relation $v_R >> v_L$) could contribute to the gauge bosons masses. 

The Higgs sector consists of only a pair of left-right symmetric isodoublets 
$\Phi_L (2,1,1) \oplus \Phi_R (1,2,1)$ with the following Higgs potential 
\begin{eqnarray}
\mathcal{V} &=& - \left(\mu^2_L\, \Phi^\dagger_L\, \Phi_L + 
                        \mu^2_R\, \Phi^\dagger_R\, \Phi_R\right) \nonumber \\
            &+&   \frac{\rho_1}{2}\, \bigg[ (\Phi^\dagger_L \Phi_L)^2 + (\Phi^\dagger_R \Phi_R)^2 
            \bigg] \nonumber \\
            &+&   \rho_2\, \left(\Phi^\dagger_L\, \Phi_L\right) 
                               \left(\Phi^\dagger_R\, \Phi_R \right)
\end{eqnarray}
The minimum of the potential corresponds to
$\left\langle \Phi_{L}\right\rangle= v_L/ \sqrt{2}, \quad
\langle \Phi_{R} \rangle= v_R/\sqrt{2}$. Choosing $\mu_R \geq \mu_L$ guarantees 
$v_R \geq v_L$. In the unitary gauge, there are two physical Higgs bosons: 
$h_L\equiv \text{Re} \Phi^0_L$ and $h_R \equiv \text{Re} \Phi_R$, these two states, 
in principle could mix with each other with mixing angle $\theta_\phi \simeq 
\left(\frac{\rho_2}{\rho_1}\right) \left(\frac{v_L}{v_R}\right)$ for $v_R \gg v_L$. 
Their masses are given by 
$$M^2_{h_R}\simeq \rho_1\, v^2_R, \quad M^2_{h_L}\simeq \rho_1 
\left(1-\frac{\rho^2_2}{\rho^2_1} \right) v^2_R $$

\subsection{Gauge boson mass}
From Eq. (\ref{eq:completelagrangian}), we can see that the relevant gauge boson 
mass terms as follow:
\begin{eqnarray}
L_{\rm boson}&=&\left|\left(-g\frac{1}{2}\tau.W_{\mu L}-g'\frac{Y}{2}B_{\mu}\right) 
\Phi_{L}\right|^2 \nonumber \\
&+&\left|\left(-g\frac{1}{2}\tau.W_{\mu R}-g'\frac{Y}{2}B_{\mu}\right) 
\Phi_{R}\right|^2
 \label{eq:boson1}
\end{eqnarray}
After substituting the vacuum expectation values of the Higgs fields: 
\begin{eqnarray}
\left\langle \Phi_{L}\right\rangle=\bordermatrix{&\cr
&0\cr
&v_L\cr},\;\left\langle \Phi_{R}\right\rangle=\bordermatrix{&\cr
&0\cr
&v_R\cr},
 \label{eq:vevhiggs}
\end{eqnarray} 
into the relation (\ref{eq:boson1}), we obtain 
\begin{eqnarray}
L_{\rm boson}&=&\frac{g^2\, v^2_L}{4}\left\{\left(W_{\mu L}^{1}\right)^2 
+\left(W_{\mu L}^{2}\right)^2\right\} \nonumber \\
&+&\frac{v^2_L}{4}\left(g\, W_{\mu L}^{3}- g'\, B_{\mu}\right)^2 \nonumber\\
&+&\frac{g^2\, v^2_R}{4}\left\{\left(W_{\mu R}^{1}\right)^2 
+\left(W_{\mu R}^{2}\right)^2\right\} \nonumber \\
&+&\frac{v^2_R}{4}\left(g\, W_{\mu R}^{3}-g'\, B_{\mu}\right)^2
\label{eq:lagrangian2}
\end{eqnarray}
Let us define:
\begin{eqnarray}
W_{\alpha}^{\pm}=\frac{1}{\sqrt{2}}\left(W_{\mu \alpha}^{1}\mp i\, W_{\mu \alpha}^{2}\right), 
Z_{\mu \alpha}=\frac{gW_{\mu \alpha}^{3}-g'B_{\mu \alpha}}{\sqrt{g^2+g'^2}},\\ 
A_{\mu \alpha}=\frac{g'W_{\mu \alpha}^{3}+gB_{\mu \alpha}}{\sqrt{g^2+g'^2}}, 
Z_{\mu \alpha}^{'}=W_{\mu \alpha}^{3},\ \ \ \ \ \ \ \ \ \ \ 
 \label{eq:boson}
\end{eqnarray}
where $\alpha=L,R$. With this definition, the gauge boson mass can read 
from eq.(\ref{eq:lagrangian2}) as:
\begin{eqnarray}
L_{\rm boson}&=&M_{W_{L}}^2\, W_{L}^{+}W_{L}^{-}
          +M_{W_{R}}^2\, W_{R}^{+}W_{R}^{-} \nonumber \\ 
&+&M_{Z_{L}}^2\, Z_{\mu L}Z_{L}^{\mu}+M_{Z_{R}}^2\, Z_{\mu R}Z_{R}^{\mu} 
\nonumber \\
&+&M_{A}^2\, A_{\mu}A^{\mu}
 \label{eq:bosonmass1}
\end{eqnarray}
where the respective masses appear in the above Lagrangian are given below
\begin{eqnarray}
& &M_{W_L}=\frac{g\,v_L}{2}, M_{W_R}=\frac{g\, v_R}{2}, M_{A}=0, \nonumber \\
& &
M_{Z_{L}}=\frac{v_L\sqrt{g^2+g'^2}}{2}, M_{Z_{R}}=\frac{v_R\sqrt{g^2+g'^2}}{2}
 \label{eq:bosonmass}
\end{eqnarray}
\subsection{Fermion mass}
We shall discuss here how fermion masses arise in this particular approach. The key idea 
of the model is to suppose the existence of weak iso-singlets heavy fermions in one-to-one 
correspondence with the light ones. In order to generate the masses of the usual SM fermions, 
we introduce some heavy charged singlets to construct the Yukawa couplings to the 
Higgs and fermion doublets so that we can derive the SM Yukawa couplings by integrating 
out these singlets (see the Ref. \cite{Berezhiani:1983hm, Berezhiani:1985in, Berezhiani:1991ds, Berezhiani:1992pj,Gu:2010yf}). 
These heavy isosinglet vector like fermions includes: 
color triplet with electric charge $+2/3$ as $U_{L,R}$, color triplet with electric charge 
$-1/3$ as $D_{L,R}$ and color singlet with electric charge $-1$ as $E_{L,R}$ and with 
these extra fields, the Yukawa terms can be written as 
\begin{eqnarray}
\label{lagrangian1a} \mathcal{L}&\supset&
-y_D^{}\left(\bar{q}_L^{}\Phi_L^{}D_R^{}+\bar{q}_R^{}\Phi_R^{}D_L^{}\right)-M_D^{}\bar{D}_L^{}D_R^{}\nonumber\\
&&-y_U^{}\left(\bar{q}_L^{}\tilde{\Phi}_L^{}U_R^{}+\bar{q}_R^{}\tilde{\Phi}_R^{}U_L^{}\right)-M_U^{}\bar{U}_L^{}U_R^{}\nonumber\\
&&-y_E^{}\left(\bar{l}_L^{}\Phi_L^{}E_R^{}+\bar{l}_R^{}\Phi_R^{}E_L^{}\right)-M_E^{}\bar{E}_L^{}E_R^{}\nonumber\\
&&+\textrm{H.c.}\nonumber \\
&\Rightarrow&
-y_d^{}\bar{q}_L^{}\Phi_L^{}d_R^{}-y_u^{}\bar{q}_L^{}\tilde{\Phi}_L^{}u_R^{}-y_e^{}\bar{l}_L^{}\Phi_L^{}e_R^{}\nonumber\\
&&+\textrm{H.c.}\,,
\end{eqnarray}
where the SM Yukawa couplings are given by
\begin{subequations}
\begin{eqnarray}
\label{smyd}
y_d^{}&=&-y_D^{L}\frac{v_R^{}}{M_D^{}}y_D^{R\dagger}\,,\\
\label{smyu}
y_u^{}&=&-y_U^{L}\frac{v_R^{}}{M_U^{}}y_U^{R\dagger}\,,\\
\label{smye} y_e^{}&=&-y_E^{L}\frac{v_R^{}}{M_E^{}}y_E^{R\dagger}\,.
\end{eqnarray}
\end{subequations}
Here we have chosen the base where the mass matrices $M_{D,U,E}^{}$
are real and diagonal.

In the neutrino sector, we consider the left- and right-handed
neutral singlets $S_{L,R}$ with the Yukawa couplings and the masses as below,
\begin{eqnarray}
\label{lagrangian1b}
\mathcal{L}&\supset&-y_S^{}\left(\bar{\ell}_L^{}\tilde{\Phi}_L^{}S_R^{}
+\bar{\ell}_R^{}\tilde{\Phi}_R^{}S_L^{}\right)
-M_S^{D}\bar{S}_L^{}S_R^{}\nonumber\\
&&-\frac{1}{2}M_S^M \left( \bar{S}_L^c S_L^{}+ \bar{S}_R^c
S_R^{}\right)+\textrm{H.c.}\,.
\end{eqnarray}
At this stage, we do not want the Yukawa couplings $\bar{\ell}_L^{}\tilde{\Phi}_L^{}S_L^{c}$, 
$\bar{\ell}_R^{}\tilde{\Phi}_R^{}S_R^c$ and their CP conjugates. This can be 
achieved by imposing a discrete symmetry as well as global and local symmetries. For example, let us
consider a $U(1)_X^{}$ local symmetry under which $D_{L,R}^{},U_{L,R}^c, 
E_{L,R}^{},S_{L,R}^c,\Phi_{L,R}^{\ast}$ carry a quantum number $X=1$. Clearly, 
this $U(1)_X^{}$ is free of gauge anomaly. In this context, the Yukawa couplings 
and the Dirac mass terms in Eqs. (\ref{lagrangian1a}) and (\ref{lagrangian1b}) are
allowed while the Majorana mass terms in Eq. (\ref{lagrangian1b}) are forbidden. 
To break this $U(1)_X^{}$, we can introduce a singlet scalar $\eta$ with Yukawa 
couplings to the neutral singlets $S_{L,R}^{}$, 
\begin{eqnarray}
\mathcal{L}&\supset& -\frac{1}{2}f_S^{} \left( \eta\, \bar{S}_L^c
S_L^{}+\eta^\ast_{} \bar{S}_R^c S_R^{}\right)+\textrm{H.c.}\,.
\end{eqnarray}
Through the above Yukawa interactions, the Majorana masses in Eq.
(\ref{lagrangian1b}) can be given by
\begin{eqnarray}
M_S^M=f_S^{} \langle \eta \rangle\,.
\end{eqnarray}
By integrating out the neutral singlets, the full neutrino masses
would contain a Dirac mass term and two Majorana ones,
\begin{eqnarray}
\label{lagrangian2} \mathcal{L}\supset-\frac{1}{2}\overline{\nu}_L^{}
M_L \nu_L^c-\frac{1}{2}\overline{N}_R^{} M_R^{}
N_R^c-\overline{\nu}_L^{} M_D^{} N_R^{}+\textrm{H.c.}
\end{eqnarray}%
with
\begin{subequations}
\begin{eqnarray}
\label{massl}
M_L^{}&=&-y_S^{}\frac{1}{M_S^M}y_S^T v_L^2\,,\\
\label{massr}
M_R^{}&=&-y_S^{}\frac{1}{M_S^M}y_S^T v_R^2\,,\\
\label{massd} M_D^{}&=&y_S^{}\frac{1}{M_S^M}(M_S^{D})^T_{}
\frac{1}{M_S^M}y_S^\dagger v_L^{}v_R^{}\,.
\end{eqnarray}
\end{subequations}
Here we have assumed
\begin{eqnarray}
\label{assumption1} M_S^M\gg M_S^D,y_S^{}v_R^{},y_S^{}v_L^{}\,,
\end{eqnarray}
by choosing the base where the Majorana mass matrix $M_N^M$ is real
and diagonal,
\begin{eqnarray}
M_S^M=\textrm{diag}\{M_1^{},M_2^{},M_3^{}\}\simeq M.
\end{eqnarray}
Clearly, the right-handed neutrinos will give their left-handed
partners an additional Majorana mass term through the seesaw since
their Dirac masses are not vanishing. This contribution is indeed negligible,
\begin{eqnarray}
\delta
M_L^{}=-M_D^{}\frac{1}{M_R^{\dagger}}M_D^T=\mathcal{O}\left[\left(\frac{M_S^D}{M_S^M}\right)^2_{}\right]M_L^{}\ll
M_L^{}\,.
\end{eqnarray}
Therefore, we can well define the left- and right-handed Majorana
neutrinos,
\begin{subequations}
\begin{eqnarray}
\nu &=&\nu_L^{}+\nu_L^c\,,\\
N &=&N_R^{}+N_R^c\,,
\end{eqnarray}
\end{subequations}
Diagonalization of the light neutrino mass matrix $m_\nu =M_L$, through 
lepton flavour mixing matrix $U_{\rm PMNS}$ \cite{Maki:1962mu} gives us three 
light Majorana neutrinos $m^{\rm diag}_{\rm light}= U_{\rm PMNS}\, M_L\, 
U^T_{\rm PMNS}=\textrm{diag} \{m_1, m_2, m_3 \}$. If we look the structure 
of light neutrino mass matrix $M_L$ and heavy neutrino mass matrix $M_N$, 
then it is clear that both the matrix can be simultaneously diagonalized by 
the same unitary matrix $U_{\rm PMNS}$, i.e  $M^{\rm diag}_{\rm heavy}= 
U_{\rm PMNS}\, M_N\, U^T_{\rm PMNS}\, \frac{v^2_R}{v^2_L} = \textrm{diag} 
\{M_1, M_2, M_3 \}$. Hence, one can correlate the eigenvalues of the light and 
heavy Majorana neutrino which in turn gives $m_\nu \propto M_N$.

In other words, one can write the light left-handed and heavy right-handed Majorana 
neutrinos mass matrices in terms of the diagonal eigenvalues of light neutrinos 
as
\begin{eqnarray*}
m_\nu=M_L&=&U^\dagger_{\rm PMNS}\textrm{diag} \{m_1^{},m_2^{},m_3^{}\}U^\ast_{\rm PMNS}\,,\\
M_N=M_R&=&U^\dagger_{\rm PMNS}\textrm{diag}\{m_1^{},m_2^{},m_3^{}\}U^\ast_{\rm PMNS}\frac{v_R^2}{v_L^2}\,.
\end{eqnarray*}
%
where $m_1$, $m_2$ and $m_3$ are the absolute masses of light Majorana neutrinos 
and are chosen to be real.

\begin{figure}[htb!]
\includegraphics[width=9cm,height=3.8cm]{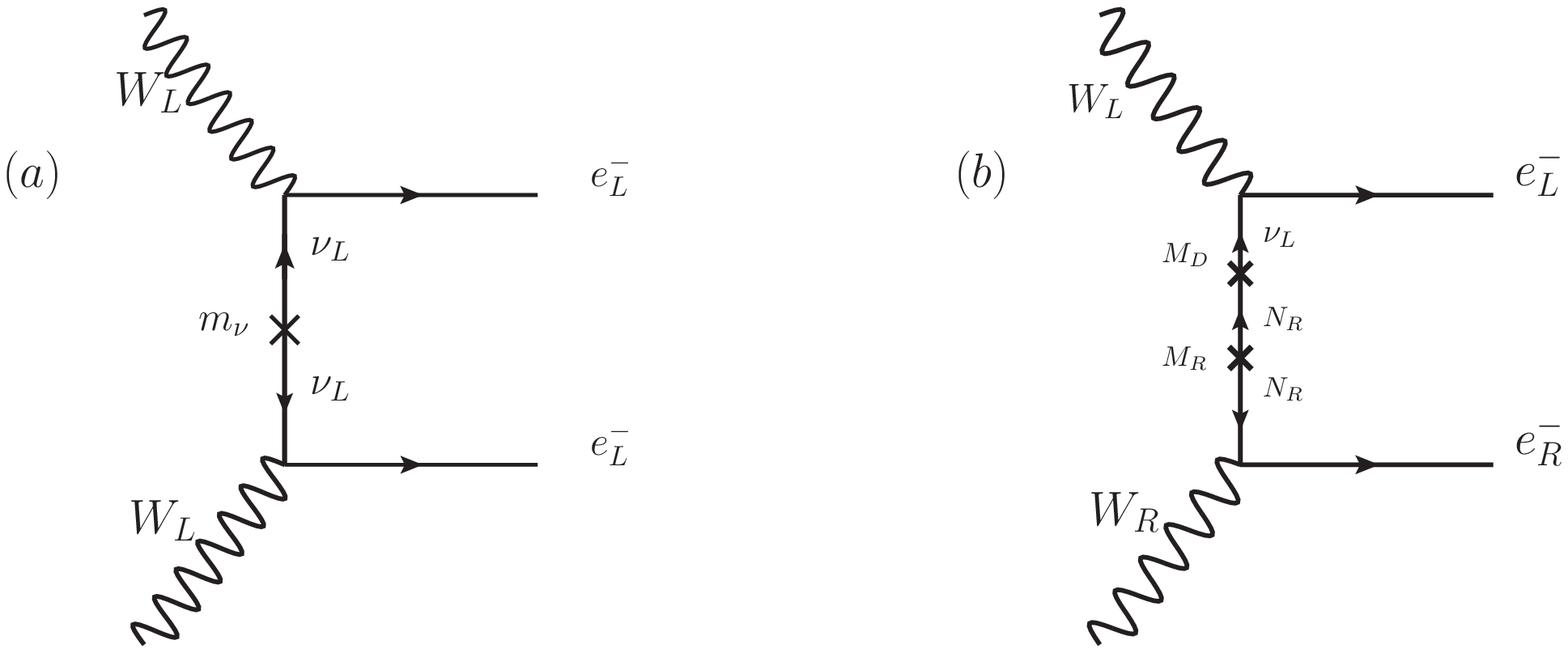}
\end{figure}
\begin{figure}[htb]
\includegraphics[width=9cm,height=3.8cm]{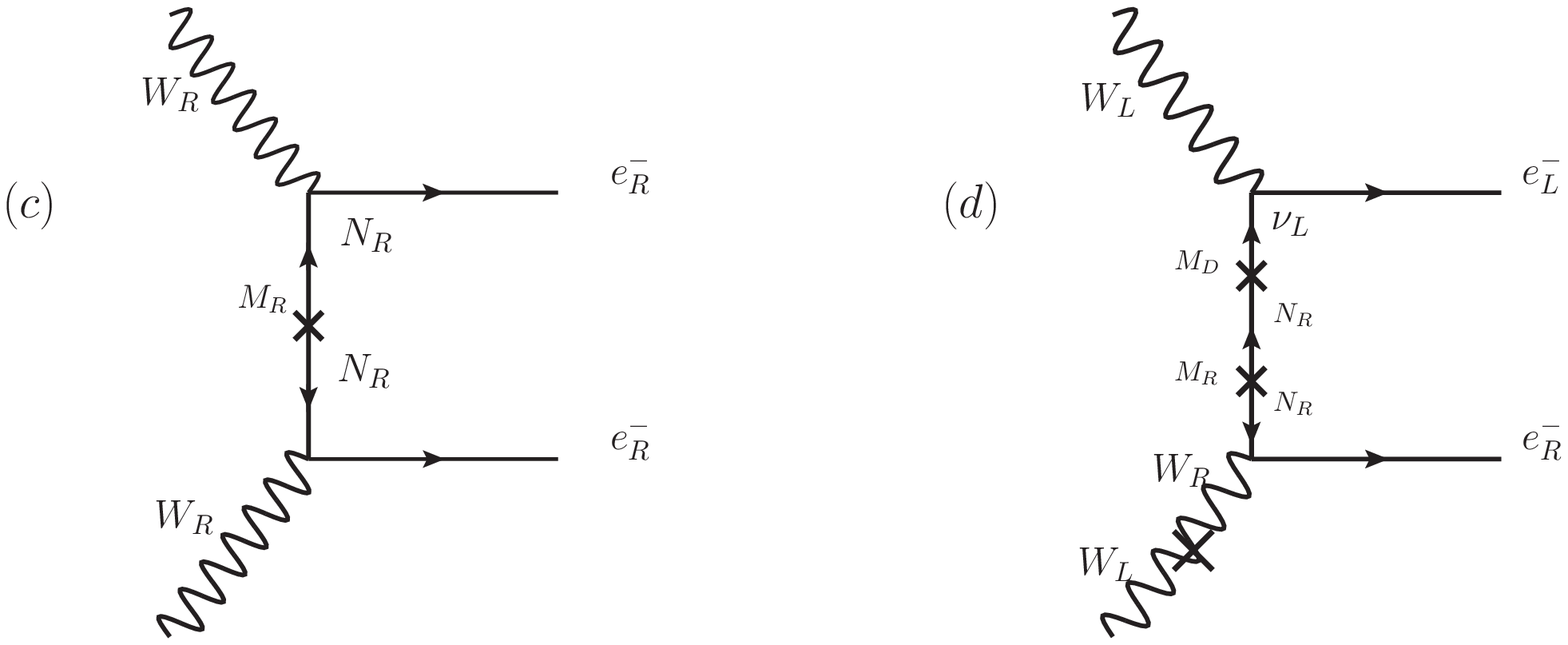}
\caption{The Feynman diagrams leading to neutrinoless double beta decay in the presence of 
right-handed current. The nucleon part that couples to $W$ bosons is omitted here, and the 
{\it Fig. \ref{fig:feyndiag-LR}(a)} is the standard process via light left-handed neutrino 
exchange {\it Fig. \ref{fig:feyndiag-LR}(b), Fig. \ref{fig:feyndiag-LR}(c), and Fig. \ref{fig:feyndiag-LR}(d)} 
are those involving right-handed current.}
\label{fig:feyndiag-LR}
\end{figure}
\section{Neutrinoless double beta decay}
In this section, we shall present the lepton number violating processes such as 
neutrinoless double beta decay in left-right symmetric model without having a 
scalar bidoublet. We shall examine how the $\nonubb$ is controlled by heavy Majorana 
neutrinos having mass around $(1-10)$ MeV. If left-right symmetry exists at high 
energy, then the contribution of the right-handed current is expected at low energy 
from the exchange of right-handed weak $W_R$ boson. The Feynman diagrams that give 
rise to neutrinoless double beta decay are depicted in Fig. \ref{fig:feyndiag-LR}(a), \ref{fig:feyndiag-LR}(b), 
\ref{fig:feyndiag-LR}(c), and \ref{fig:feyndiag-LR}(d).

The corresponding Feynman amplitude for these above diagrams is depicted in 
following table as 

\begin{table}[htb]
\begin{center}
\begin{tabular}{|c|c|}
\hline
{\bf Feynman diagrams}     &  {\bf Amplitude } 
\rule{0pt}{2.1em}\\  [10pt]
\hline 
{\bf Fig. \ref{fig:feyndiag-LR}(a)}  &  
 $\mathcal{A}_{a} \propto G^2_F\, \frac{U^2_{e\,i}\, m_{\nu i}}{p^2}$
  \rule{0pt}{2.1em}\\  [10pt]
{\bf Fig. \ref{fig:feyndiag-LR}(b)}  &  
 $\mathcal{A}_{b} \propto G^2_F\, \left(\frac{M^2_{W_L}}{M^2_{W_R}}\right)\,
U^2_{e\,i}\, \left(\frac{M_D}{M_R} \right)\, \frac{1}{|p|}$
  \rule{0pt}{2.1em}\\  [10pt]
{\bf Fig. \ref{fig:feyndiag-LR}(c)}  &  
 $\mathcal{A}_c \propto G^2_F\, 
 \left(\frac{M^4_{W_L}}{M^4_{W_R}}\right)\,U^2_{e\,i}\, 
\frac{M_{R\, i}}{p^2}$
  \rule{0pt}{2.1em}\\  [10pt]
{\bf Fig. \ref{fig:feyndiag-LR}(d)}  &  
 $\mathcal{A}_{d} \propto G^2_F\,\left(\frac{M^2_{W_L}}{M^2_{W_R}}\right)\,
 U^2_{e\,i}\, \left(\frac{M_D}{M_R} \right)\, 
\zeta_{L-R}\, \frac{1}{|p|}
 $
  \rule{0pt}{2.1em}\\  [10pt]
\hline
\end{tabular}
\caption{ABLE I. Analytic formulas for amplitudes for different Feynman diagrams 
      in neutrinoless double beta decay process as described in the text.}
\end{center}
\end{table}
In this table, $G_F=1.2 \times 10^{-5} \text{GeV}^{-2}$ is the Fermi constant, $M_{W_R}$ is the 
right-handed charged gauge boson mass, $\zeta_{L-R}$ is the $W_L-W_R$ mixing and $p^2$ is the 
neutrino virtuality. In order to estimate the relative contributions of different terms, 
it is worth to note here that we shall analyze the effect of neutrinoless double beta decay 
while the representative set of parameters in this model are:  $M_{W_R} \sim$ 10 TeV and 
the  heaviest right-handed neutrino mass around $\sim (1- 10)$ MeV. With this set of parameters, 
the relevant dominant contributions are found to be
\begin{eqnarray} 
& &\mathcal{A}_{a} \propto \frac{G^2_F}{p^2}\, \left(U^2_{e\,i}\, m_{\nu i}\right)
        \sim \frac{G^2_F}{p^2} \times 10^{-2} \text{eV} \nonumber \\
& &\mathcal{A}_{c} \propto \frac{G^2_F}{p^2}\, 10^{-8}\, 10^{7} \text{eV}
       \sim \frac{G^2_F}{p^2} \times 10^{-1} \text{eV} \nonumber
\end{eqnarray}

\subsection{The standard contribution from left-handed current}
\begin{figure}[htb]
\includegraphics[width=7cm,height=5cm]{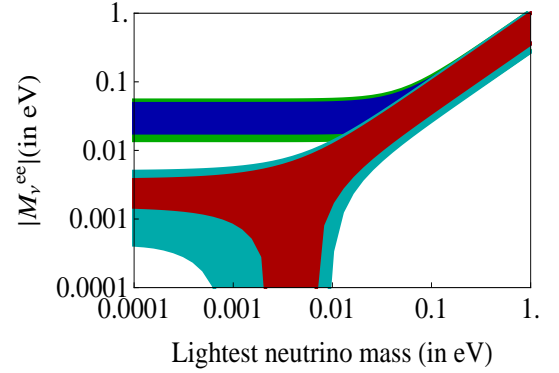}
\caption{The generic contribution from light neutrino mass with $\theta_{13}$ 
from \cite{An:2012eh} to the neutrinoless double beta decay. Here the top (blue and 
green) horizontal bands are for inverted neutrino mass hierarchy, and the bottom 
(red and cyan) bands are for hierarchical neutrino masses.}
\label{fig:lr-light}
\end{figure}

In generic contribution to total decay width for neutrinoless double beta decay 
$(0\nu\, \beta \beta)$, which comes from the left-handed light neutrinos as 
exchange particle, is given as, 
\begin{equation}
\Gamma_{0\nu } = G^{0\nu}\, 
\left|\frac{\mathcal{M}^{0\nu}}{m_e}\right|^2      
|M_\nu^{ee}|^2  ,
\end{equation}
where $G^{0\nu}$ is a phase space factors, $m_e$ is the electron mass, 
${\mathcal{M}^{0\nu}}$ is the nuclear matrix element and the effective 
Majorana mass is given by
\begin{equation} 
|M_\nu^{ee}| = |U^2_{ej} \, m_j|, 
\label{effectiveml}
\end{equation} 
Here $U_{ej}$ are the elements of the lepton mixing matrix $U_{\rm PMNS}$ given in 
\cite{Maki:1962mu} which contains three mixing angles and three phases (one Dirac and two 
Majorana phases). It is worth to emphasize here that the neutrinoless double beta decay 
experiment can probe the phases which crucially depends on the pattern of the neutrino 
masses i.e whether neutrinos are Normal, or, Inverted, or, quasi-degenerate and on the 
magnitude of the neutrino masses. One can parametrize the effective Majorana mass in terms 
of the elements of $U_{\rm PMNS}$ and mass eigenvalues as
\begin{eqnarray} 
|M_\nu^{ee}| &=& \big| \cos_{12}^2\, \cos_{13}^2\, m_1 + e^{ 2 i \alpha_2}\, \sin_{12}^2\, \cos_{13}^2\, m_2 
\nonumber \\ &+& e^{2 i \alpha_3}\, \sin_{13}^2\, m_3 \big|. 
\label{mnuee-effective}
\end{eqnarray}
This contribution of effective Majorana mass is depicted in Fig. \ref{fig:lr-light} which gives the value of 
the effective Majorana mass as a function of lightest neutrino mass. To generate the required 
plot, we have used the 3-$\sigma$ ranges and the best-fit values of the mass squared differences 
and mixing angles $\sin^2 \theta_{12}$, $\sin^2 \theta_{23}$ from global analysis of oscillation 
data \cite{Fogli:2011qn} and value of $\sin^2 \theta_{13}$ from the recent measurement of DayaBay 
experiment \cite{An:2012eh}. In particular, the representative values of the parameters which has been taken 
in this model, in order to give the result shown in Fig. \ref{fig:lr-light}, are as follows
\begin{eqnarray}
& &\Delta m^2_{\rm {sol}} [10^{-5} \mbox{eV}^2] \quad \quad \text{7.58} [\text{best-fit}]\quad \text{6.99-8.18} [\text{3-sigma}] \nonumber \\ 
& &|\Delta m^2_{\rm {atm}}| [10^{-3} \mbox{eV}^2]| \quad \text{2.35} [\text{best-fit}] \quad \text{2.06-2.67} [\text{3-sigma}]\nonumber \\ 
& &\sin^2\theta_{12} \quad \quad \quad \text{0.306} [\text{best-fit}] \quad \text{0.259-0.359} [\text{3-sigma}] \nonumber \\ 
& &\sin^2\theta_{23} \quad \quad \quad \text{0.42} [\text{best-fit}]\quad ~~\text{0.34-0.64} [\text{3-sigma}]\nonumber \\ 
& &\sin^2\theta_{13} \quad \quad \quad \text{0.023} [\text{best-fit}] \quad \text{0.009 - 0.037} [\text{3-sigma}]\nonumber
\end{eqnarray}
In the plot, we need to explain how effective Majorana mass probes which kind of mass pattern 
of neutrinos. As shown in Fig. \ref{fig:lr-light}, the cyan band for NH corresponds to varying the parameters 
in their 3$\sigma$ range whereas the red band corresponds to the best-fit parameters where the 
$\sin^2 \theta_{12}$ values are taken from recent Daya-Bay result. In both figures the Majorana 
phases are varied between 0 to 2$\pi$. In the same manner, The green band for IH corresponds 
to varying the parameters in their 3$\sigma$ range whereas the blue band corresponds to the 
best-fit parameters. We will not present the detailed analysis of this figure since this has 
already been discussed elaborately in ref. \cite{Chakrabortty:2012mh}. We shall now move to next subsection 
where the dominant contribution comes from the right-handed current and present an analysis for 
the result obtained with MeV mass range of RH Majorana neutrinos. 

\subsection{New contribution from right-handed current}
From the discussion of the light and heavy Majorana neutrino masses which is stated in the end 
of section-II, it is found that they are related with each other as $m_{j} \propto M_{j}$, 
where the proportionality 
factor is $v^2_R/v^2_L$. Before relating heavy RH neutrinos in terms of light neutrino masses, we 
will first present the different hierarchy pattern of the light neutrinos as follows
\begin{itemize}
\item In case of normal scheme (NH), the light neutrino masses $m_2$ and $m_3$ can be expressed in terms 
of the lightest light neutrino mass $m_1$ as
$$m_2 = \sqrt{m_1^2 +\Delta m_{\rm sol}^2},\quad m_3 = \sqrt{m_1^2 +\Delta m_{\rm atm}^2 + \Delta m_{\rm sol}^2}$$
and their mass hierarchy is $ m_1 < m_2 << m_3$.
\item Inverted hierarchy (IH) implies $m_3 << m_1 \sim m_2$ and the light neutrino masses $m_1$ and $m_2$ 
can be written in terms of the lightest light neutrino mass, which is $m_3$ in this case, as
$$m_1 = \sqrt{m_3^2 +\Delta m_{\rm atm}^2}, \quad m_2 = \sqrt{m_3^2 +\Delta m_{\rm sol}^2 +\Delta m_{\rm atm}^2 }$$
\item The quasi-degenerate limit correspond to $m_1 \approx m_2 \approx m_3 >> \sqrt{\Delta m^2_{atm}}$.
\end{itemize}
In the following, we will present the relation between heavy right-handed neutrino masses in terms 
of light left-handed neutrinos for various mass spectra and try to analyze the behavior of effective 
Majorana mass $M^{ee}_{N}$ as a function of lightest light left-handed neutrinos. 

\subsubsection*{\bf Hierarchical pattern of the neutrino masses}
It is important mention here that the value of $M_{W_R}$ has to be at least 10 TeV in order 
to get MeV scale of heaviest right-handed (RH) neutrino mass so that the new contributions to neutrinoless 
double beta decay coming from right-handed current can be comparable. In presenting the 
analytical behavior of the neutrinoless double beta decay contribution coming from the 
right-handed current, one should first give the heavy RH neutrino mass ratios to those 
of light neutrinos which are given below 
\begin{eqnarray}
\frac{M_{_1}}{M_{_3}} = \frac{m_{1}}{m_{3}}, \quad
\text{and} \quad \frac{M_{2}}{M_{3}} = \frac{m_{2}}{m_{3}}. \nonumber
\label{eqn:Mne} 
\end{eqnarray}
where the value of the heaviest RH neutrino mass $M_3$ is fixed around MeV range. With this input, the 
expression for $M^{ee}_N$ is given by
\begin{eqnarray}
& &|M^{ee}_N|_{NH} = \left( \frac{M_{W_L}}{M_{W_R}} \right)^4\, \sum_{j}\, U^2_{ej}\, M_{j} \nonumber \\
&&\hspace{0.11cm}=\left( \frac{M_{W_L}}{M_{W_R}} \right)^4\,M_{3}\, \bigg| 
\cos^2 \theta_{12}\, \cos^2 \theta_{13}\, \frac{m_1}{m_3} \nonumber \\
&&\hspace{0.1cm}+ \sin^2 \theta_{12} \cos^2 \theta_{13} e^{2 i \alpha_2}\,\frac{m_2}{m_3}
           + \sin^2 \theta_{13}\,e^{2 i \alpha_3} \bigg|  
\label{eqn:M} 
\end{eqnarray}

\begin{figure}[htb]
\includegraphics[width=7cm,height=5cm]{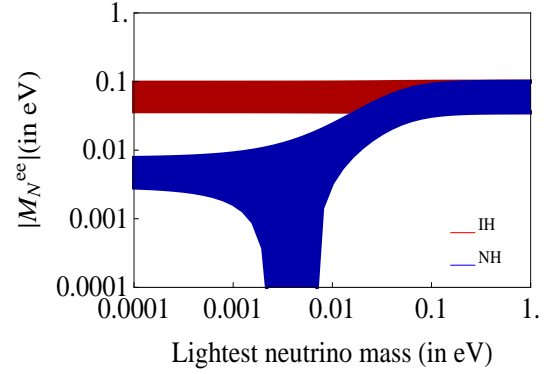}
\caption{The new dominant contribution to neutrinoless double beta decay coming 
from the right-handed current having $M_j$ around MeV and right-handed $W$-bosons around 
10 TeV. Here the upper (red) band is for inverted hierarchical and the lower (blue) band is for 
          hierarchical light neutrino masses.}
\label{fig:lr-WR}
\end{figure}

In the purely hierarchical case, $10^{-5} \text{eV}< m_1 < 10^{-3} \text{eV}$, one can 
write $m_2 \simeq \sqrt{\Delta m_{\rm sol}^2}$, $m_3 \simeq \sqrt{\Delta m_{\rm atm}^2}$. Given the input 
parameters in our model, the ratio between left- and right-handed charged gauge boson masses 
is found to be $10^{-8}$, the ratio between solar and atmospheric mass square difference is 
$m_2/m_3 \simeq \sqrt{\Delta m_{\rm sol}^2 /\Delta m_{\rm atm}^2} = \{0.16, 0.2 \}$ corresponds to 
minimum and maximum value respectively. Since $m_1$ is very small, the first term in eqn. 
(\ref{eqn:M}) gives negligible contribution and hence can be neglected. With the choice made 
for $M_3$ at 5 MeV scale, the effective Majorana mass is
\begin{eqnarray}
\label{eqn:M-small}
 |M^{ee}_N|_{NH}&=&0.05\bigg|\sin^2 \theta_{12} \cos^2 \theta_{13}\, 
           \sqrt{\frac{\Delta m_{\rm sol}^2}{\Delta m_{\rm atm}^2 }}\,  e^{2 i \alpha_2} \nonumber \\
           &+& \sin^2 \theta_{13}\, e^{2 i \alpha_3} \bigg|
\end{eqnarray}
The maximum and minimum values of $|M^{ee}_N|_{NH}$ corresponds to the phase values 
$\alpha_2,~ \alpha_3=0, \pi$ and $\alpha=0, \pi; \alpha_3=\pi/2$ respectively. As 
shown in Fig. \ref{fig:lr-WR}, the blue band in the regime $10^{-5} \text{eV}< m_1 < 10^{-3} \text{eV}$ 
corresponds to minimum and maximum values as follows
\begin{displaymath} 
\left\{ 
\begin{array}{l} 
|M^{ee}_N|_{NH} (\text{max}) = 0.0075 \\ 
|M^{ee}_N|_{NH} (\text{min}) = 0.0055
\end{array} \right. 
\end{displaymath} 
For intermediate hierarchical values of $m_1$, says $10^{-3} \text{eV}< m_1 < 10^{-2} \text{eV}$, still 
the first term in eqn. (\ref{eqn:M}) can be neglected. For illustration, one can see that the first term 
of eqn. (\ref{eqn:M-small}) is small because the smallness of $ \sqrt{\Delta m_{\rm sol}^2 /\Delta m_{\rm atm}^2}$, 
at the same time, the second term is also suppressed due to the factor $\sin^2 \theta_{13}$. As a result, there 
is cancellation occurs in this regime due to relative phase cancellation of $\alpha_2$ and $\alpha_3$.

\subsubsection*{\bf Inverted Hierarchy of the neutrino masses}

In this case, the other heavy RH neutrino masses can be expressed in terms 
of light neutrino masses (keeping $M_2$ fixed which is the heaviest RH neutrino 
mass) as
\begin{eqnarray}
\frac{ M_{1}}{ M_{2}} = \frac{m_{1}}{m_{2}},\quad \text{and} \quad 
\frac{M_{3}}{M_2} = \frac{m_{3}}{m_{2}}  .
              \nonumber 
\end{eqnarray}
Now the expression for $M^{ee}_N$ becomes
\begin{eqnarray}
\label{eqn:ih}
& &|M^{ee}_N|_{IH} = \left( \frac{M_{W_L}}{M_{W_R}} \right)^4\,M_{2} \bigg|
\cos^2 \theta_{12}\, \cos^2 \theta_{13}\,\frac{m_1}{m_2} \nonumber \\
& &\hspace{0.1cm}+\sin^2 \theta_{12} \cos^2 \theta_{13} e^{2 i \alpha_2}
           + \sin^2 \theta_{13}\,e^{2 i \alpha_3}\,\frac{m_1 }{m_2}\bigg | 
\end{eqnarray}
Before illustrating the analytical behavior of this contribution, it should be noted 
here that the value of $m_3$ in the case inverted hierarchy is such that $m_3 \ll \sqrt{\Delta m_{\rm sol}^2}$, 
$m_1 \simeq \sqrt{\Delta m_{\rm atm}^2}$ and $m_2 \simeq \sqrt{\Delta m_{\rm sol}^2}$. 
Since the factor $ m_3 /m_2$ is very small in this regime 
and the value of $\sin^2 \theta_{13}$ is also very small, the last term of the eqn. (\ref{eqn:ih}) 
can be safely neglected. Now the effective Majorana mass in this inverted hierarchical scheme is 
given below
\begin{eqnarray}
\label{eqn:ih-fix}
|M^{ee}_N|_{IH} 
&=& 0.05 \bigg|
\cos^2 \theta_{12}\, \cos^2 \theta_{13}\,\sqrt{\Delta m_{\rm atm}^2 /\Delta m_{\rm sol}^2} \nonumber \\
&+&\sin^2 \theta_{12} \cos^2 \theta_{13} e^{2 i \alpha_2} \bigg | 
\end{eqnarray}
Similarly, the same arguments discussed in above subsection will gives the maximum 
and minimum values of the $|M^{ee}_N|_{IH}$ as
\begin{displaymath} 
\left\{ 
\begin{array}{l} 
|M^{ee}_N|_{IH} (\text{max}) = 0.1 \\ 
|M^{ee}_N|_{IH} (\text{min}) = 0.05
\end{array} \right. 
\end{displaymath}

\subsubsection*{\bf Quasi-degenerate pattern of the neutrino masses}

In this limit, $m_1 \sim m_2 \sim m_3 \sim m_0$ which implies 
$$m_0 \gg \sqrt{\Delta m^2_{\rm sol}}, \sqrt{\Delta m^2_{\rm atm}}$$
Since quasi-degeneracy pattern of light neutrino masses also implies 
quasi-degeneracy in heavy right-handed neutrino sector which implies 
$$M_1 \approx M_2 \approx M_3 =M_0$$
where $M_0$ is common absolute mass of heavy RH neutrinos which is at 
MeV scale. In this situation, one can have the relation for the heavy 
neutrino contribution to the effective mass is 
\begin{eqnarray}
|M^{ee}_N|_{QD} &=& \left( \frac{M_{W_L}}{M_{W_R}} \right)^4\, M_{0} \bigg|
\cos^2 \theta_{12}\, \cos^2 \theta_{13} \nonumber \\
&+&\sin^2 \theta_{12} \cos^2 \theta_{13} e^{2 i \alpha_2}
           + \sin^2 \theta_{13}\,e^{2 i \alpha_3}\bigg | \nonumber 
\end{eqnarray}
From this relation, we can conclude that effective neutrino mass from RH current 
is independent of lightest neutrino mass in the quasi-degenerate limit. In other 
words, the value of $|M^{ee}_N|$ remains constant with increasing $m_1$.

\subsection{Total contribution}
The total dominant contribution to neutrinoless double beta decay in left-right 
model, in which the scalar sector consists of two isodoublets $\Phi_L$ and $\Phi_R$ without 
having bidoublet, is given by
\begin{equation}
\Gamma_{0\nu }= G^{0\nu} \cdot \left| \frac{\mathcal{M}^{0\nu}}{m_e}\right|^2   
|m_{\rm eff}^{\rm ee}|^2
\end{equation}

\begin{figure}[htb]
\includegraphics[width=7cm,height=5cm]{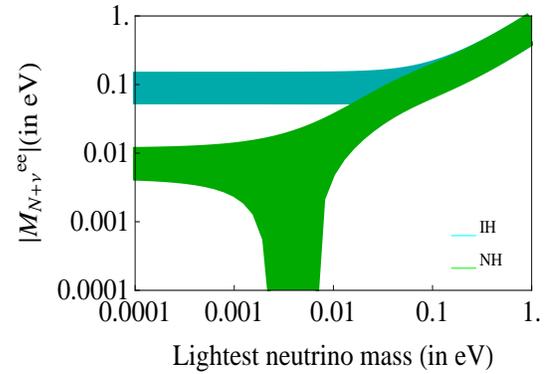}
\caption{The total contribution to neutrinoless double beta decay in 
left-right models without having a scalar bidoublet. Here the upper (cyan) band and the lower (green) band correspond to inverted hierarchy 
          and normal hierarchy of the light neutrino masses respectively.}
\label{fig:lr-combine}
\end{figure}

The effective neutrino mass contribution to neutrinoless double beta decay is
\begin{eqnarray}
|m_{\rm eff}^{\rm ee}|^2& = &\Bigg( |  U^2_{ej} \, m_j\ |^2 + \Bigg| \frac{M^4_{W_L}}{M_{W_R}^4} 
                          U_{\,ej}^2\, M_{N_j}   \Bigg|^2 \Bigg)\,, \nonumber \\
                      & = & |  M_{\nu}^{ee} |^2 + | M^{ee}_N |^2 \  
\label{eqn:Mnee_nldbd+lr}    
\end{eqnarray}
where the individual contribution are $M_{\nu}^{ee} = U^2_{ej} \, m_j$ and $M^{ee}_N =   
\frac{M^4_{W_L}}{M_{W_R}^4}\, U_{\,ej}^2\, M_{N_j}$. This combine contribution is illustrated 
in Fig. \ref{fig:lr-combine}.
\section{Comments on Cosmological constraints}
We shall discuss in this section whether the MeV scale RH neutrinos for $M_{W_R}$ lying in 1-10 TeV region 
is consistent with the big-bang nucleosynthesis (BBN) bound and from the over closing of the Universe. We are in a problematic situation 
when $M_{W_R}$ lies around TeV scale, which in turn gives over-abundance of RH neutrinos $N$, because the 
$SU(2)_R$ gauge interaction keep them in thermal equilibrium when the temperature is high. Also, 
if RH neutrinos are allowed to decay later than
late after the BBN era, they end up destroying the abundance of light elements, which in turns gives 
$\tau_N \lesssim \text{sec}$, that translates into a lower bound on $M_N$. 

Let us first consider the case of the {\em heavy} regime, with $M_N \gtrsim m_\pi + m_\ell$, where $N$ decays 
sufficiently fast into a charged (anti)lepton and a pion with the following decay rate
\begin{equation}
\begin{split}
  \Gamma_{N \to \ell \pi} \!=\! \frac{G_F^2 |V_{ud}^{qR}|^2 |V_{\ell N}^R|^2 f_\pi^2\, M_N^3}{8 \pi}
  \frac{M_W^4}{M_{W_R}^4} \bigl[ \left(1 - x_\ell^2 \right)^2 - 
  \\
  x_\pi^2 \left(1 + x_\ell^2 \right) \bigr] \!
  \left[ \! \left(1 \! - \! \left(x_\pi + x_\ell \right)^2 \right) \left(1\! - \! \left( x_\pi - x_\ell \right)^2 \right) \! \right]^\frac12,
\end{split}
\label{eqNDecayRate}
\end{equation}
where $x_{\pi,\ell} = m_{\pi, \ell} / M_N$, $ V^R$ is the right-handed lepton mixing matrix,
$V^{qR}$ is the analog quark one and $f_{\pi} = 130\, \text{MeV}$ is the pion decay constant.  We recall that
$V_{ud}^{qR}\simeq V_{ud}^{qL}\simeq 0.97$; on the other hand, the leptonic mixing involved depends
on the mass hierarchy and on the flavor of the charged lepton into which the RH neutrino is
decaying.  As one can check from~\eqref{eqNDecayRate}, for $M_N> m_\pi+m_\ell$, the above process
guarantees that $\tau_N$ is safely shorter than a second. Hence the constraints coming from the cosmology 
gives $M_N > 140$ MeV. This range of RH neutrino mass will push up the $M_{W_R}$ scale beyond TeV scale which 
spoils the possible probe of our scenario in near future like at LHC.

The prescribed scenario discussed above suffers from serious problem when $M_N$ lies below $< 140$ MeV, 
the life-time becomes longer than a second, a decaying $N$ would pump too much entropy into the universe. 
The point is that they decouple relativistically at the temperature
\begin{equation}
  T_D^N= T_D^\nu \left(\frac{M_{W_R}}{M_{W}}\right)^{\frac43},
\end{equation}
where $T_D^\nu\simeq1\, \text{MeV}$ is the neutrino decoupling temperature. Therefore, for a representative
value of $M_{W_R}\sim 5\,\text{TeV}$,
\begin{equation}
  T_D^N\simeq 250\,\text{MeV}\,.
\end{equation}
Then, since between $T_D^N$ and 1 MeV, only muons and pions decouple, at BBN $N$'s are almost
equally abundant as light neutrinos. The only way out would be to make $N$ stable and to avoid the
over-closure of the universe, lighter than about eV~\cite{Seljak:2006bg, Fogli:2008ig}. As a result, 
we are in a scenario where extra species are contributing to BBN.  Actually,
this situation seems to be preferred and a recent study suggests~\cite{Hamann:2010bk, Hamann:2011ge} that four
light neutrinos give the best fit to cosmological data, while five is disfavored and six is
basically excluded.

\section{Conclusion}
We have discussed neutrinoless double beta decay in the context of left-right symmetric models 
with the minimal Higgs content, which is different from standard version of $L-R$ model that make 
use of $L-$ and $R-$ Higgs triplets and a Higgs bidoublet for the fermion mass generation. The 
scalar sector model consists of two Higgs doublets $\Phi_L$ and $\Phi_R$ without invoking triplets 
and bi-doublet, and the fermion masses are generated by integrating out the extra vector-like 
heavy quarks and leptons. 
In gauge sector, there is no mixing between left- and right-handed weak gauge bosons 
at tree level, but can induced at one loop level. In this particular scenario where 
the light neutrino and heavy Majorana neutrino are related with each other and diagonalized 
by the same Pontecorvo-Maki-Nakagawa-Sakata (PMNS) matrix, the neutrinoless double beta decay receives important contributions 
from the right handed current. In fact, the choice we made for the right handed Majorana 
neutrino mass around MeV range in the context of neutrinoless double beta decay, 
the last formula of Sec-II predicts
the right-handed gauge boson mass $M_{W_R}$ to be at least order of 5 TeV.

\section{Acknowledgment } 
The author would like to thank Prof. Z.G. Berezhiani for useful comments and Goran Senjanovic, 
and Miha Nemevsek for useful discussions on neutrinoless double beta decay. The author also 
acknowledges support and facilities from the Institute of Physics, Bhubaneswar where a part 
of this work was carried out.

\bibliographystyle{apsrev}
\bibliography{ref}
\end{document}